\documentclass[runningheads,a4paper]{llncs}

\usepackage{color}
\usepackage{graphicx}
\usepackage{booktabs}
\usepackage{subfigure}
\usepackage{listings}
\usepackage{multirow}
\usepackage{mathtools}
\usepackage{amsmath}
\usepackage[]{algorithm2e}

\begin{document}

\title{Predicting Transportation Modes of GPS Trajectories using Feature Engineering and Noise Removal}
\author{Mohammad Etemad\inst{1}%https://orcid.org/0000-0002-3770-180X
\and Am\'ilcar Soares J\'unior\inst{1} \and Stan Matwin\inst{1}\inst{2}
}
\titlerunning{Predicting Trans. Mode using Feature Engineering\&Noise Removal}  
\institute{
Institute for Big Data Analytics, Dalhousie University, Halifax
\and
Institute for Computer Science, Polish Academy of Sciences, Warsaw}

\maketitle            % typeset the title of the contribution

\begin{abstract}
Understanding transportation mode  from GPS (Global Positioning System) traces is an essential topic in the data mobility domain. 
In this paper, a framework is proposed to predict transportation modes. 
This framework follows a sequence of five steps: (i) data preparation, where GPS points are grouped in trajectory samples; (ii) point features generation; (iii) trajectory features extraction; (iv) noise removal; (v) normalization. 
We show that the extraction of the new point features: bearing rate,  the rate of rate of change of the bearing rate and the global and local trajectory features, like medians and percentiles enables many classifiers to achieve high accuracy (96.5\%) and f1 (96.3\%) scores.
We also show that the noise removal task affects the performance of all the models tested.
Finally, the empirical tests where we compare this work against state-of-art transportation mode prediction strategies show that our framework is competitive and outperforms most of them. 

\end{abstract}

\section{Introduction}

Research on trajectory analysis is a mature area since positioning devices are now used to track people, vehicles, vessels, and animals. 
In the case of trajectory data, the object's movement is represented as a discrete collection of spatiotemporal points.

A domain where trajectories are frequently analyzed is the prediction of transportation modes from users, which is essential for cities and people to reduce travel time and traffic congestion. 
Transportation mode estimation involves two steps \cite{zheng2008understanding}: (i) extraction of segments of the same transportation modes; and  (ii) classification of transportation modes for each segment.
For the first step, several segmentation algorithms have been proposed in the past years and include temporal-based \cite{Stenneth2011}, cost function-based \cite{soaresjunior2015} and semantic-based methods \cite{spaccapietra2008}.
For the second step, which is the focus of this work, the classification (or prediction) of the transportation modes is performed by creating domain expert features for supervised classification (e.g., the distance between consecutive points, velocities, acceleration, and bearing).

We classify the research in transportation modes prediction regarding the type of features in two branches: (i) domain expert features; and (ii) learned features.
From raw GPS data points (e.g., latitude, longitude and time) it is possible to calculate many attributes regarding the moving object's movement. 
Examples include distance traveled between points, estimated speed, bearing, acceleration, etc. 
For segments of trajectories, it is possible to extract mean, median, minimum, maximum, standard deviations, etc., of point-wise features.
These are examples of domain expert features employed to predict transportation modes.
Examples of works that apply domain expert features include \cite{lin2014mining,zheng2008understanding}.

In this work, we also explore the effects of noise removal in the prediction of transportation modes. 
Dealing with noise in trajectories is essential because GPS recorder devices are not accurate in the moving object's positioning due to many reasons like satellite geometry, signal blockage, atmospheric conditions, and receiver design features/quality. 
By removing GPS noise, it is expected that the derived features from the trajectories are more likely to represent the standard pattern of a transportation mode. 

Noise-perturbed GPS data influences the quality of the domain expert features, e.g.  distance traveled, speed or acceleration are susceptible to errors. 
It is important to point out that these errors may impact the distributions of values, where statistics like the mean, in trajectory segments of transportation modes. 
This uncertainty of data can lead a classifier to create models that are not able to accurately predict a transportation mode from a trajectory.
Thus, the works in transportation mode prediction are classified regarding the (i) presence or (ii) absence of noise removal strategies. 
An example of work in the transportation mode prediction that does not deal with noise removal is \cite{zheng2008understanding}. 
In others, like \cite{Yanyun2017CNN,jiang2017trajectorynet,dabiri2018inferring,endo2016deep,xiao2017identifying}, noise is removed. 
This paper applies domain expert features and noise removal to predict transportation are as follows: (i) we introduce new point and trajectory features; (ii) we propose a framework composed of 5 steps for transportation mode prediction; (iii) we compare the proposed approach with state-of-art strategies and show that our results are competitive.

%The remainder of the paper is organized as follows. 
%Our framework to predict transportation modes is detailed in Section \ref{sec:model}. 
%The experimental setup and results of this work are fully described and analyzed in Section \ref{sec:experiments}.
%Finally, Section \ref{sec:conclusions} shows the conclusions and future works. 

%\input{./sections/framework.tex}
\section{A framework for transportation mode prediction}
\label{sec:model}

\begin{figure}[ht]
\caption{The steps of the proposed framework to predict transportation modes}
\centering
\includegraphics[width=0.8\textwidth]{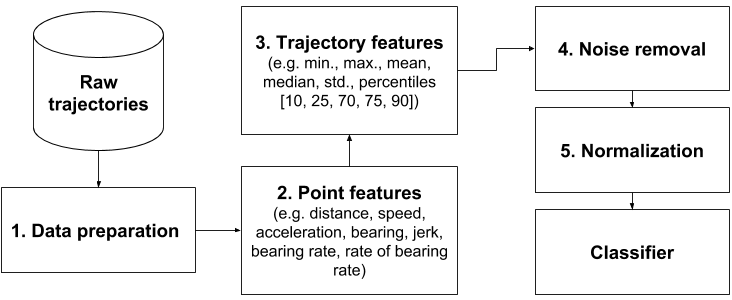}
\label{fig:model}
\end{figure}

In this section, we present the sequence of steps used in this work to predict transportation modes (Figure \ref{fig:model}).
This framework has five steps and is described in detail below.

In this work, we define a trajectory as a sequence of GPS points that belongs to the same transportation mode. 
In the first (step 1), we group the raw GPS points by $user id$, $day$ and transportation mode to create trajectory samples. 
We discard trajectory samples with less than 10 GPS points because these examples may affect our model since trajectories with low quality may be created. 

In this work, we calculate some point features (step 2) that were used previously in literature \cite{zheng2008understanding}: distance, speed, acceleration, jerk\cite{dabiri2018inferring}, and bearing. 

Two new features are introduced in this work, named bearing rate, and the rate of bearing rate. They are detailed as follows.
The bearing rate was computed using Eq. \ref{eq:5}, where $B_i$ and $B_{i+1}$ are the bearing values in points $i$ and $i+1$, and $\Delta t$ is the time difference. 

\begin{equation}
\label{eq:5}
B_{rate(i+1)}=(B_{i+1}-B_{i})/\Delta t
\end{equation}

Some moving objects tend to change the bearing more often because they commute in a straightforward route. This behavior can be captured by using the rate of the bearing rate. 
This feature is calculated using Eq.\ref{eq:6}.

\begin{equation}
\label{eq:6}
Br_{rate(i+1)} =(B_{rate(i+1)}-B_{rate(i)})/\Delta t
\end{equation}

After calculating all the point features for each trajectory, we extract some statistical attributes referred to as trajectory features (step 3). 
Trajectory features are divided into two different types: (i) global trajectory features, which summarize information regarding the whole trajectory in a single value; and (ii) local trajectory features, which describe a local part of the trajectory. 
In this work, we extracted global features like the Minimum, Maximum, Mean, Median, and Standard Deviation values of each trajectory point feature to feed our classifier. 
The local trajectory features extracted in this work was the percentiles of every point feature.
Five different percentiles were extracted (10, 25, 50, 75, and 90) and were used in the models tested in this work. 
In summary, we compute 70 trajectory features (10 statistical measures including five global and five local features calculated for 7 point features) for each transportation mode example. 

In step 4, the framework deals with noise in the data. 
In this work, we used a simple method called median filter to create a mask.  The method is described in Algorithm \ref{eq:40} ($threshold=3$) and it removes the noise based on $speed mean$ (i.e. the average speed of a trajectory) attribute since a human can classify the transportation mode mostly by knowing the mean speed of a trajectory. 

\begin{algorithm}
\label{eq:40}
 \KwData{Speed mean of trajectories}
 \KwResult{mask vector to remove the noisy trajectories }
 $difference \longleftarrow |{speed mean}_{Trajectory} - median(speed {mean})|$\;
 ${median}\_{difference} \longleftarrow median(difference)$\;
 
  \eIf{median\_difference == 0}{
   $indicator\longleftarrow0$\;
   }{
   $indicator\longleftarrow difference/median\_difference$\;
  
 }
 \KwRet{indicator $>$ threshold}\;
 \caption{mask the noisy samples to remove from dataset using median}
\end{algorithm}

Finally, we normalized the features (step 5) using the Min-Max normalization method, since this method preserves the relationship between the values to transform features to the same range and improve the quality of classification process \cite{han2011data}.

\section{Experiments}
\label{sec:experiments}

In this section, we detail the experiments performed in this work to validate our framework. 
The data used in this work is the GeoLife GPS dataset, that was collected by Microsoft Research Asia from April 2007 to October 2011 \cite{zheng2008understanding}. 
The dataset has a 5,504,363 number of records labeled by eleven transportation modes: taxi (4.41\%); car (9.40\%); train (10.19\%); subway (5.68\%); walk (29.35\%); airplane (0.16\%); boat (0.06\%); bike (17.34\%); run (0.03\%); motorcycle (0.006\%); and bus (23.33\%).

In the literature, we observed different sub-selections of these classes for evaluating transportation mode prediction strategies;  
therefore, we decided to select different target subsets for comparing our result with other papers.

To evaluate the performance of classifiers in this work we used the Accuracy and the F1 measure.
In all our experiments, we used a 10-fold cross-validation strategy and computed a paired t-test to verify if the difference in the means were statistically different. 
We executed our framework with different classifiers such as Decision Tree (DT) (with $max depth$ equals five), Random Forest (RF) (with 50 trees estimators), Neural Network (NN), Naive Bayes (NB), and Quadratic Discriminant Analysis (QDA). In all cases, the random forest surpasses all the other classifiers in both accuracy and f1.

Subsequently, we compared the RF using all the steps of our framework against the results of five papers. 
It is important to point out that all these papers  reported their accuracy values on the Geolife dataset. 
Table \ref{tbl:cmp} shows a side-by-side comparison between some related works and the results of our framework. 
Our work does not surpass Jiang's et al. accuracy \cite{jiang2017trajectorynet} but outperforms all the others.
It is important to highlight that the complexity and high training time of the RNN model used in his work may not be worth the 1.42\% difference in accuracy. 

\begin{table}[ht]
\centering
\caption{Comparison of accuracy and f1 measure of proposed model against related works}
\label{tbl:cmp}
\begin{tabular}{|l|c|c|c|l|}
\hline
\multicolumn{2}{|c|}{\textbf{Related work}} & \multicolumn{3}{c|}{\textbf{Proposed Model}}       \\ \hline
\textbf{Reference: classes used in the experiments}            & \textbf{acc}           & \textbf{acc}        & \multicolumn{2}{c|}{\textbf{f1}} \\ \hline
Dabiri et al. \cite{dabiri2018inferring}          : walk, bike, bus, driving, and train&84.8\%&\textbf{93.35\%} & \multicolumn{2}{c|}{93.22\%} \\ \hline
Jiang et al.\cite{jiang2017trajectorynet}: bike, car, walk, and bus&\textbf{97.9}\%&96.45\% & \multicolumn{2}{c|}{96.31\%} \\ \hline
 Xiao et al. \cite{xiao2017identifying}
 : walk, bus\&taxi, bike, car, subway, and train&90.77\%&\textbf{93.19\%} & \multicolumn{2}{c|}{92.81\%} \\ \hline
Zheng et al.\cite{zheng2008understanding}       
 : walk, driving, bus, and bike&76.2\%&\textbf{93.61\%} & \multicolumn{2}{c|}{93.51\%} \\ \hline
Endo et al.\cite{endo2016deep}             
 : walk, car, taxi, bike, subway, bus, and train&83.2\%&\textbf{90.20\%} & \multicolumn{2}{c|}{89.95\%} \\ \hline
\end{tabular}
\end{table}

Finally, we evaluated the effects of noise removal performed by our framework. 
We established as a baseline the performance of our framework using the data to train classifiers with noise and without noise (clean).
Table \ref{tbl:f1} shows the mean of the f1 values obtained by 10-fold cross-validation for the different group of classes.
We can observe in Table \ref{tbl:f1} that for all classifiers and different subgroups of classes, performance gains ranging from 2.56 (Decision Tree, using classes of \cite{endo2016deep}) to 28.15 (QDA, using classes of \cite{zheng2008understanding}) in f1.

\begin{table}[h]
\centering
\caption{F1 measures to classifiers for different class groups.}
\label{tbl:f1}
\begin{tabular}{|l|l|l|l|l|l|l|l|l|l|l|}
\hline
\multirow{2}{*}{\textbf{\begin{tabular}[c]{@{}c@{}}Reference\end{tabular}}} & \multicolumn{2}{c|}{\textbf{DT}}                                                                                                                                           & \multicolumn{2}{c|}{\textbf{RF}}                                                                                                                                           & \multicolumn{2}{c|}{\textbf{NN}}                                                                                                                                           & \multicolumn{2}{c|}{\textbf{NB}}                                                                                                                                           & \multicolumn{2}{c|}{\textbf{QDA}}                                                                                                                                               \\ \cline{2-11} 
                                                                                 & \multicolumn{1}{c|}{\textbf{\begin{tabular}[c]{@{}c@{}}with\\ noise\end{tabular}}} & \multicolumn{1}{c|}{\textbf{\begin{tabular}[c]{@{}c@{}}clean\end{tabular}}} & \multicolumn{1}{c|}{\textbf{\begin{tabular}[c]{@{}c@{}}with\\ noise\end{tabular}}} & \multicolumn{1}{c|}{\textbf{\begin{tabular}[c]{@{}c@{}}clean\end{tabular}}} & \multicolumn{1}{c|}{\textbf{\begin{tabular}[c]{@{}c@{}}with\\ noise\end{tabular}}} & \multicolumn{1}{c|}{\textbf{\begin{tabular}[c]{@{}c@{}}clean\end{tabular}}} & \multicolumn{1}{c|}{\textbf{\begin{tabular}[c]{@{}c@{}}with\\ noise\end{tabular}}} & \multicolumn{1}{c|}{\textbf{\begin{tabular}[c]{@{}c@{}}clean\end{tabular}}} & \multicolumn{1}{c|}{\textbf{\begin{tabular}[c]{@{}c@{}}with\\ noise\end{tabular}}} & \multicolumn{1}{c|}{\textbf{\begin{tabular}[c]{@{}c@{}}clean\end{tabular}}} \\ \hline

\textbf{Dabiri et al. \cite{dabiri2018inferring}}&\multicolumn{1}{c|}{85.56}& \multicolumn{1}{c|}{92.31}& \multicolumn{1}{c|}{88.07}& \multicolumn{1}{c|}{93.22}& \multicolumn{1}{c|}{85.18}& \multicolumn{1}{c|}{89.87}& \multicolumn{1}{c|}{63.30}& \multicolumn{1}{c|}{82.91}& \multicolumn{1}{c|}{54.76}& \multicolumn{1}{c|}{79.83}\\ \hline

\textbf{Jiang et al.\cite{jiang2017trajectorynet}}&\multicolumn{1}{c|}{88.26}& \multicolumn{1}{c|}{95.47}& \multicolumn{1}{c|}{91.56}& \multicolumn{1}{c|}{96.31}& \multicolumn{1}{c|}{88.63}& \multicolumn{1}{c|}{94.11}& \multicolumn{1}{c|}{65.68}& \multicolumn{1}{c|}{85.19}& \multicolumn{1}{c|}{54.70}& \multicolumn{1}{c|}{82.55}\\ \hline

\textbf{Xiao et al. \cite{xiao2017identifying}}&\multicolumn{1}{c|}{84.38}& \multicolumn{1}{c|}{89.79}& \multicolumn{1}{c|}{88.75}& \multicolumn{1}{c|}{92.81}& \multicolumn{1}{c|}{82.93}& \multicolumn{1}{c|}{89.01}& \multicolumn{1}{c|}{51.40}& \multicolumn{1}{c|}{70.03}& \multicolumn{1}{c|}{47.81}& \multicolumn{1}{c|}{71.45}\\ \hline

\textbf{Zheng et al.\cite{zheng2008understanding}}&\multicolumn{1}{c|}{85.62}& \multicolumn{1}{c|}{91.92}& \multicolumn{1}{c|}{88.72}& \multicolumn{1}{c|}{93.51}& \multicolumn{1}{c|}{85.76}& \multicolumn{1}{c|}{91.33}& \multicolumn{1}{c|}{64.61}& \multicolumn{1}{c|}{84.22}& \multicolumn{1}{c|}{51.33}& \multicolumn{1}{c|}{79.48}\\ \hline

\textbf{Endo et al.\cite{endo2016deep}}&\multicolumn{1}{c|}{79.53}& \multicolumn{1}{c|}{82.09}& \multicolumn{1}{c|}{85.57}& \multicolumn{1}{c|}{89.95}& \multicolumn{1}{c|}{79.33}& \multicolumn{1}{c|}{85.70}& \multicolumn{1}{c|}{57.31}& \multicolumn{1}{c|}{72.68}& \multicolumn{1}{c|}{49.13}& \multicolumn{1}{c|}{72.30}\\ \hline

\end{tabular}
\end{table}

Finally, Table \ref{tbl:acc} shows the mean of the accuracy values obtained by 10-fold cross-validation. 
For all classifiers and different subgroups of classes and classifiers, performance gains ranging from 3.36 (Decision Tree, using classes of \cite{endo2016deep}) to 29.04 (QDA, using classes of  \cite{jiang2017trajectorynet}) in accuracy were observed. 
The results presented in this section indicate that dealing with noise in transportation mode prediction is an important topic, and the lack of this step in the classification task decreases the performance of the classifiers.

\begin{table}[ht]
\centering
\caption{Accuracy to classifiers for different class groups.}
\label{tbl:acc}
\begin{tabular}{|l|l|l|l|l|l|l|l|l|l|l|}
\hline
\multirow{2}{*}{\textbf{\begin{tabular}[c]{@{}c@{}}Class \\ group\end{tabular}}} & \multicolumn{2}{c|}{\textbf{DT}}                                                                                                                                           & \multicolumn{2}{c|}{\textbf{RF}}                                                                                                                                           & \multicolumn{2}{c|}{\textbf{NN}}                                                                                                                                           & \multicolumn{2}{c|}{\textbf{NB}}                                                                                                                                           & \multicolumn{2}{c|}{\textbf{QDA}}                                                                                                                                               \\ \cline{2-11} 
                                                                                 & \multicolumn{1}{c|}{\textbf{\begin{tabular}[c]{@{}c@{}}with\\ noise\end{tabular}}} & \multicolumn{1}{c|}{\textbf{\begin{tabular}[c]{@{}c@{}}clean\end{tabular}}} & \multicolumn{1}{c|}{\textbf{\begin{tabular}[c]{@{}c@{}}with\\ noise\end{tabular}}} & \multicolumn{1}{c|}{\textbf{\begin{tabular}[c]{@{}c@{}}clean\end{tabular}}} & \multicolumn{1}{c|}{\textbf{\begin{tabular}[c]{@{}c@{}}with\\ noise\end{tabular}}} & \multicolumn{1}{c|}{\textbf{\begin{tabular}[c]{@{}c@{}}clean\end{tabular}}} & \multicolumn{1}{c|}{\textbf{\begin{tabular}[c]{@{}c@{}}with\\ noise\end{tabular}}} & \multicolumn{1}{c|}{\textbf{\begin{tabular}[c]{@{}c@{}}clean\end{tabular}}} & \multicolumn{1}{c|}{\textbf{\begin{tabular}[c]{@{}c@{}}with\\ noise\end{tabular}}} & \multicolumn{1}{c|}{\textbf{\begin{tabular}[c]{@{}c@{}}clean\end{tabular}}} \\ \hline

\textbf{Dabiri et al. \cite{dabiri2018inferring} }&\multicolumn{1}{c|}{85.54}& \multicolumn{1}{c|}{92.36}& \multicolumn{1}{c|}{88.47}& \multicolumn{1}{c|}{93.35}& \multicolumn{1}{c|}{85.54}& \multicolumn{1}{c|}{90.13}& \multicolumn{1}{c|}{63.56}& \multicolumn{1}{c|}{83.28}& \multicolumn{1}{c|}{53.65}& \multicolumn{1}{c|}{79.76}\\ \hline

\textbf{Jiang et al.\cite{jiang2017trajectorynet}}&\multicolumn{1}{c|}{88.41}& \multicolumn{1}{c|}{95.54}& \multicolumn{1}{c|}{91.91}& \multicolumn{1}{c|}{96.45}& \multicolumn{1}{c|}{88.80}& \multicolumn{1}{c|}{94.21}& \multicolumn{1}{c|}{63.70}& \multicolumn{1}{c|}{84.31}& \multicolumn{1}{c|}{53.03}& \multicolumn{1}{c|}{82.07}\\ \hline

\textbf{Xiao et al. \cite{xiao2017identifying}}&\multicolumn{1}{c|}{85.01}& \multicolumn{1}{c|}{89.96}& \multicolumn{1}{c|}{89.33}& \multicolumn{1}{c|}{93.19}& \multicolumn{1}{c|}{83.61}& \multicolumn{1}{c|}{89.43}& \multicolumn{1}{c|}{51.96}& \multicolumn{1}{c|}{69.90}& \multicolumn{1}{c|}{46.59}& \multicolumn{1}{c|}{70.99}\\ \hline

\textbf{Zheng et al.\cite{zheng2008understanding}}&\multicolumn{1}{c|}{85.77}& \multicolumn{1}{c|}{92.13}& \multicolumn{1}{c|}{89.09}& \multicolumn{1}{c|}{93.61}& \multicolumn{1}{c|}{86.10}& \multicolumn{1}{c|}{91.45}& \multicolumn{1}{c|}{64.36}& \multicolumn{1}{c|}{84.53}& \multicolumn{1}{c|}{50.85}& \multicolumn{1}{c|}{79.50}\\ \hline

\textbf{Endo et al.\cite{endo2016deep}}&\multicolumn{1}{c|}{80.25}& \multicolumn{1}{c|}{83.61}& \multicolumn{1}{c|}{86.36}& \multicolumn{1}{c|}{90.20}& \multicolumn{1}{c|}{80.27}& \multicolumn{1}{c|}{86.28}& \multicolumn{1}{c|}{56.66}& \multicolumn{1}{c|}{73.27}& \multicolumn{1}{c|}{47.92}& \multicolumn{1}{c|}{71.60}\\ \hline

\end{tabular}
\end{table}

\section{Conclusions and Future Works}
\label{sec:conclusions}

In this work, we propose a framework for transportation mode prediction using feature engineering and noise removal. 
The results showed that the newly engineered features (e.g., bearing rate, and rate of bearing rate) and the application of a noise removal technique improve the performance of all tested classifiers. 
We intend to extend this work in two directions:
(i) test and evaluate different noise removal techniques like wavelet-based, MCMC and fast Fourier based denoising methods, and (ii) investigate the performance of trajectory segmentation algorithms and include this step in our framework.  

\subsubsection*{Acknowledgments}
The authors would like to thank NSERC (Natural Sciences and Engineering Research Council of Canada) for financial support.

\bibliographystyle{plain}
\bibliography{biblio}

\begin{thebibliography}{10}

\bibitem{dabiri2018inferring}
Sina Dabiri and Kevin Heaslip.
\newblock Inferring transportation modes from gps trajectories using a
  convolutional neural network.
\newblock {\em Transportation Research Part C: Emerging Technologies},
  86:360--371, 2018.

\bibitem{endo2016deep}
Yuki Endo, Hiroyuki Toda, Kyosuke Nishida, and Akihisa Kawanobe.
\newblock Deep feature extraction from trajectories for transportation mode
  estimation.
\newblock In {\em Pacific-Asia Conference on Knowledge Discovery and Data
  Mining}, pages 54--66. Springer, 2016.

\bibitem{han2011data}
Jiawei Han, Jian Pei, and Micheline Kamber.
\newblock {\em Data mining: concepts and techniques}.
\newblock Elsevier, 2011.

\bibitem{jiang2017trajectorynet}
X.~Jiang, E.~N. Souza, A.~Pesaranghader, B.~Hu, D.~L. Silver, and S.~Matwin.
\newblock Trajectorynet: An embedded gps trajectory representation for
  point-based classification using recurrent neural networks.
\newblock {\em arXiv preprint arXiv:1705.02636}, 2017.

\bibitem{soaresjunior2015}
A.~Soares J\'unior, B.~N. Moreno, V.~C. Times, S.~Matwin, and L.~A.~F. C.
\newblock {GRASP-UTS}: an algorithm for unsupervised trajectory segmentation.
\newblock {\em Int. J. of Geographical Information Science}, 29(1):46--68,
  2015.

\bibitem{lin2014mining}
Miao Lin and Wen-Jing Hsu.
\newblock Mining gps data for mobility patterns: A survey.
\newblock {\em Pervasive and Mobile Computing}, 12:1--16, 2014.

\bibitem{spaccapietra2008}
S.~Spaccapietra, C.~Parent, M.~L. Damiani, J.~A. Macedo, F.~Porto, and
  C.~Vangenot.
\newblock A conceptual view on trajectories.
\newblock {\em Data and Knowledge Engineering}, 65(1):126--146, 2008.

\bibitem{Stenneth2011}
Leon Stenneth, Ouri Wolfson, Philip~S. Yu, and Bo~Xu.
\newblock Transportation mode detection using mobile phones and gis
  information.
\newblock In {\em Proceedings of the 19th ACM SIGSPATIAL International
  Conference on Advances in Geographic Information Systems}, GIS '11, pages
  54--63, New York, NY, USA, 2011. ACM.

\bibitem{xiao2017identifying}
Z.~Xiao, Y.~Wang, K.~Fu, and Fan Wu.
\newblock Identifying different transportation modes from trajectory data using
  tree-based ensemble classifiers.
\newblock {\em ISPRS}, 6(2):57, 2017.

\bibitem{Yanyun2017CNN}
G.~Yanyun, Z.~Fang, C.~Shaomeng, and L.~Haiyong.
\newblock A convolutional neural networks based transportation mode
  identification algorithm.
\newblock In {\em 2017 International Conf. on Indoor Positioning and Indoor
  Navigation (IPIN)}, pages 1--7, Sept 2017.

\bibitem{zheng2008understanding}
Yu~Zheng, Quannan Li, Yukun Chen, Xing Xie, and Wei-Ying Ma.
\newblock Understanding mobility based on gps data.
\newblock In {\em Proceedings of the 10th international conference on
  Ubiquitous computing}, pages 312--321. ACM, 2008.

\end{thebibliography}

\end{document}